\documentclass[10pt,journal,cspaper,compsoc]{IEEEtran}

\usepackage{graphics,
           graphicx,
           psfrag,
           epsfig,
           amsthm,
           cite,
           amssymb,
           url,
           dsfont,
           subfigure,
           algorithm,
           algorithmic,
           balance,
           enumerate,
           color,
}
\usepackage{amsmath}

\newcommand{\sref}[1]{Section~\ref{#1}}

\newcommand{\fref}[1]{Figure~\ref{#1}}

\newcommand{\cref}[1]{Constraint~\ref{#1}}

\newcommand{\ignore}[1]{}

\IEEEoverridecommandlockouts
\begin{document}

\title{Joint Indoor Localization and Radio Map Construction with Limited Deployment Load}
\author{Sameh~Sorour,~\IEEEmembership{Member,~IEEE,}
        Yves~Lostanlen,~\IEEEmembership{Member,~IEEE,} Shahrokh~Valaee,~\IEEEmembership{Senior Member,~IEEE}
\IEEEcompsocitemizethanks{\IEEEcompsocthanksitem S. Sorour and S. Valaee are with Edward S. Rogers Sr. Department of Electrical and Computer Engineering,
    University of Toronto, 10 King's College Road, Toronto, ON, M5S 3G4, Canada,
    e-mail:\{samehsorour, valaee\}@comm.utoronto.ca.
    \IEEEcompsocthanksitem Y. Lostanlen is the Chief Technical Officer and Vice President Wireless of Siradel, France, e-mail:ylostanlen@siradel.com.
    \IEEEcompsocthanksitem This work was sponsored by Mitacs, Canada and Siradel, France.
    \IEEEcompsocthanksitem This work is an extension to our paper \cite{GC12} in Globecom 2012.}
     }

\IEEEoverridecommandlockouts

\IEEEcompsoctitleabstractindextext{
\begin{abstract}
One major bottleneck in the practical implementation of received signal strength (RSS) based indoor localization systems is the extensive deployment efforts required to construct the radio maps through fingerprinting.  In this paper, we aim to design an indoor localization scheme that can be directly employed without building a full fingerprinted radio map of the indoor environment. By accumulating the information of localized RSSs, this scheme can also simultaneously construct the radio map with limited calibration. To design this scheme, we employ a source data set that possesses the same spatial correlation of the RSSs in the indoor environment under study. The knowledge of this data set is then transferred to a limited number of calibration fingerprints and one or several RSS observations with unknown locations, in order to perform direct localization of these observations using manifold alignment. We test two different source data sets, namely a simulated radio propagation map and the environment’s plan coordinates. For moving users, we exploit the correlation of their observations to improve the localization accuracy. The online testing in two indoor environments shows that the plan coordinates achieves better results than the simulated radio maps, and a negligible degradation with 70-85$\%$ reduction in calibration load.
\end{abstract}
}

\maketitle

\section{Introduction}
Received Signal Strength (RSS)-based localization systems have attracted much attention in recent years and have been extensively studied as the most promising and relatively inexpensive solution for indoor positioning (\cite{6042868} and references therein). Their operation is mainly based on detecting and analyzing the signals of the widely deployed 802.11 access points (APs) in public indoor environments and the integrated 802.11 wireless cards in most recent mobile devices. Consequently, these systems do not require any investments in neither deploying APs nor any additional device hardware. This makes it very appealing for commercialization over other measurement based algorithms such as time-of-arrival or angle of-arrival measurements of ultra-wideband signals.

RSS-based localization techniques consist of two phases: an offline training phase (a.k.a fingerprinting phase) and an online localization phase. In the offline phase, RSS measurements are collected from all existing APs in the environment at all predefined locations. For each location, either the average of its taken measurements or all its statistics define its radio fingerprint. The collection of these fingerprints for all locations is known as the radio map. In the online localization phase, the real-time RSS samples received from the APs at the user's mobile device are compared against the stored radio map to estimate the user's current location.

As can be easily inferred from the above description, building the radio map of an indoor environment through fingerprinting can be a very exhaustive, expensive and time consuming process, especially if this environment is of large size (such as airports and giant malls). Consequently, it represents the most expensive bottleneck facing the feasible commercialization of this efficient indoor localization approach. Thus, any solutions towards reducing this deployment cost and workload is of extreme importance. Moreover, these solutions would gain even further importance if they can be used for the construction of indoor radio maps for other purposes, such as the evaluation and optimization of the wireless coverage, capacity and radio dimensioning.

Several works \cite{Bahl2000,Ji:2006:ADI:1134680.1134697,Deasy2007,Chintalapudi:2010:ILW:1859995.1860016} have tried to replace the construction of fingerprinted radio maps by other methods using indoor radio propagation model. Nonetheless, most of these propagation models cannot capture all the details of the indoor structure (e.g. exact dimensions, beams, different window heights) and dynamics (e.g. moving people, moving elevators, changing furniture locations). Consequently, all these works either achieve very unsatisfactory performance \cite{Deasy2007} or rectify the model inaccuracies through extensive calibration and exhaustive post-processing to obtain a map that can be used for more acceptable localization accuracy.

Other works have proposed algorithms to solve the problem of variations of RSS radio maps between different times or different devices using limited calibration. Examples of such algorithms are LANDMARK\cite{1192765}, LEASE \cite{1356987}, LEMT \cite{4359003}, LeManCor \cite{Pan2007} and LuMA \cite{4724994}. Despite the great improvement in cost and complexity of adapting radio maps to temporal and device variations, all these techniques still require a complete and accurate deployment radio map. Moreover, the accumulation of information adaptation over several cycles may increase the localization error eventually. Some of them suffer from dense calibration requirements or extensive processing. Finally, the adaptation process is done to generate a new map for future localization and cannot be used to directly localize targets.

\ignore{
In a previous work, we proposed an method to directly localize users without building a full fingerprinted map, using a combination of simulated radio model, obtained from a radio propagation simulator, and a limited number of calibration fingerprints. The proposed technique indeed reduces the load of fingerprinting significantly and achieves comparable results. Nonetheless, generating the radio model requires both significant material and structure knowledge from the environment as well as the locations and heights of all APs involved in the localization process. All these data should be then traced in specific softwares and fed to the simulator to generate the model. This process still represents a deployment load, which maybe less costly than the fingerprinting load, but is still not reduced enough.
}

In this paper, we propose an indoor localization solution that can be simply deployed using limited calibration and can directly localize users without building a full radio map of the indoor environment. The accumulation of this localization information can then be used to jointly construct the radio map for other purposes. The idea behind this proposed scheme is to exploit the inherent spatial correlation of RSS measurements to reduce the amount of fingerprints and perform direct localization without a full radio map. It is well known that neighboring positions usually have highly correlated radio fingerprints. If we could find a data set that can reflect this spatial correlation pattern, and knowing the actual RSSs at a few locations in the environment through fingerprinting, we can locate real-time collected RSSs via transfer learning. This localization can be done directly using manifold alignment \cite{Ham2004}, which is a semi-supervised, dimensionality reduction transfer learning method. This method learns the location of one or several RSS observations by aligning the spatial correlation information with the calibration and observation measurements in a low-dimensional space.

One data set that can represent this radio spatial correlation pattern is the simulated map using indoor radio propagation models. These maps indeed reflect, to a good extent, the radio propagation effects in space, such as power decay reflections, diffractions and fading. However, due to model inaccuracies, simulated radio maps usually suffer from neighborhood correlation outliers (i.e. non-neighboring positions having quite similar simulated RSSs). Another data set that could be used the plan coordinates. This simple data set does not need any effort to generate but yet it captures the neighborhood correlation of the physical environment in its most exact form. However, the plan coordinates does not fully reflect the radio propagation effects. In this paper, we consider both data sets to determine the effect that has the more important role in achieving a better accuracy.

For walking users, we can achieve a better performance by exploiting the correlation of their subsequent reported observations to improve the localization accuracy. This is done by performing the localization for several subsequent RSS observations simultaneously, in which we both enforce a neighborhood property between its subsequent observations when aligning the manifolds and reject localization outliers.

Whether using simulated radio maps or plan coordinates, it is obvious that the effort and requirements to deploy our proposed solution is much less compared to full fingerprinting. For it to directly operate, it only requires much fewer fingerprints along with the knowledge of the floor plan and maybe few simulations. In addition to simple deployment, \emph{our approach is easily adaptable to changes in time, device or floor plans}. In the first two cases, only few new calibrations at different times or with different devices are enough to adapt our algorithm with the same spatial correlation pattern. In case of drastic floor plan changes, the spatial correlation function of the environment can be easily adapted to it and the system can re-operate in few minutes.

The rest of the paper is organized as follows. In \sref{sec:related}, we summarize the related works to our problem.\ignore{\sref{sec:description} introduces the formal description of the localization problem.}  \sref{sec:manifold-alignment} introduces the basics of locally linear embedding and manifold alignment. We then present our proposed joint localization and radio map construction solutions using simulated radio maps and plan coordinates in \sref{sec:SRM-algorithm} and \sref{sec:PC-algorithm}, respectively. We then present the localization testing environments, data collection and results in \sref{sec:sim-L}, and those of the radio map construction in \sref{sec:sim-R}. \sref{sec:conclusion} concludes the paper.

\section{Related Work}\label{sec:related}

In \cite{1192765,1356987}, LANDMARK and LEASE proposed an adaptive offset of the RSS variations using deployed reference sniffers. This approach adapts the radio map to environmental dynamics using real-time samples, but still requires an initial map to start with. Moreover, it has been shown to be successful only with densely distributed reference sniffers. \cite{4359003} proposed LEMT that learns the functional relationship between the initial map and real-time readings (again from deployed sniffers) using nonlinear regression analysis and model trees, then applies nearest neighbor based method to find locations. LEMT requires less reference sniffers than LANDMARK and LEASE and can achieve a more effective accommodation of RSS variation. However, LEMT requires extensive processing after each RSS sniffing period by building a huge number of trees in each of them.

In \cite{Pan2007,4724994} and LeManCoR and LuMA were proposed, respectively, to transfer knowledge across different times and devices, using multiview learning and manifold alignment, respectively. Nonetheless, both algorithms are only for radio map update and still require a complete and accurate (i.e. fingerprinted) deployment radio map to start with. This update may even lose accuracy for big changes in the floor plan or if this update is repeated for several cycles. Moreover, the adaptation process is done to generate a new map for future localization and cannot be used to directly localize targets. These problems are all solved in our proposed solution.

\section{Manifold Alignment}\label{sec:manifold-alignment}
Manifolds alignment \cite{Ham2004} is a dimensionality reduction based semi-supervised transfer learning scheme. It learns mappings between a source data set and target data set that are characterized by the same underlying manifold and discovers the corresponding relationship in a low-dimensional space. The discovered correlation is used to transfer the knowledge from the source data to target data. Transfer learning using manifold alignment can be employed if two main assumptions are satisfied in the data sets to which it would be applied. The first assumption is the strength of neighborhood correlation. In other words, each of the two data sets should have a stronger correlation for neighboring data points than for further points. The second assumption is that the two data sets should possess a common lower dimensional correlation, even if they exhibit different distributions or shapes in a higher dimensional space.

In the next sections, we will briefly illustrate the mechanisms and formulation of the manifold alignment problem.

\subsection{Neighborhood Weights}
Manifold alignment is based on aligning two data sets in a lower dimension space using their correlation in that space and preserving the neighborhood correlation. Several dimensionality reduction techniques have been studied in the literature. In our work, we select the locally linear embedding technique \cite{Saul2000} as it strongly preserves neighborhood correlation in the lower dimensional space\ignore{, which fits our problem}.

In locally linear embedding (LLE), a low-dimensional embedding is constructed using a weighted graph that captures local structure in the data. For each higher dimensional data point $\mathbf{z}^{(i)}$, the $N$ data points having the smallest distance to it form its neighbor set $\mathcal{N}(i)$. Let $\left[\mathbf{z}^{\left(\mathcal{N}(i,1)\right)},\dots,\mathbf{z}^{\left(\mathcal{N}(i,N)\right)}\right]$ be the set of these $N$ neighboring data points of point $i$. We compute neighborhood weights of $\mathbf{z}^{(i)}$ using the following optimization problem:
\begin{align}\label{eq:weights-formulation}
&\arg\min_{W_{ij}} \left\{\left|\mathbf{z}^{(i)} - \sum_{j\in\mathcal{N}(i)} W_{ij}\:\mathbf{z}^{\left(\mathcal{N}(i,j)\right)}\right|^2\right\} \nonumber \\
&\quad \mbox{s.t.}\quad \sum_{j\in\mathcal{N}(i)}W_{ij}= 1\;.
\end{align}
Clearly, the closer the point $\mathbf{z}^{\left(\mathcal{N}(i,j)\right)}$ to $\mathbf{z}^{(i)}$, the higher the weight $W_{ij}$. For points $j$ not in $\mathcal{N}(i)$, the value of $W_{ij} = 0$. This optimization can be solved using a closed form solution \cite{Saul2000} as follows. Define the distance matrix $\mathbf{D}_i$ of point $i$ as:
\begin{equation}
\mathbf{D}_i=
\begin{bmatrix}
\mathbf{z}^{(i)} - \mathbf{z}^{\left(\mathcal{N}(i,1)\right)}\\
\mathbf{z}^{(i)} - \mathbf{z}^{\left(\mathcal{N}(i,2)\right)}\\
\vdots\\
\mathbf{z}^{(i)} - \mathbf{z}^{\left(\mathcal{N}(i,N)\right)}\\
\end{bmatrix}\;.
\end{equation}
Thus, the weight $W_{ij}$ between $\mathbf{z}^{(i)}$ and each neighbor $\mathbf{z}^{\left(\mathcal{N}(i,j)\right)}$ can be expressed as:
\begin{equation}\label{eq:weights-solution}
W_{ij} = \frac{\sum_{k =1}^N \left\{\left(\mathbf{D}_i\mathbf{D}_i^T\right)^{-1}\right\}_{jk}}{\sum_{m=1}^N\sum_{n=1}^N \left\{\left(\mathbf{D}_i\mathbf{D}_i^T\right)^{-1}\right\}_{mn}}\;.
\end{equation}
where $\left\{\left(\mathbf{D}_i\mathbf{D}_i^T\right)^{-1}\right\}_{uv}$ is the element on the $u$-th row and $v$-th column in the inverse of matrix $\left(\mathbf{D}_i\mathbf{D}_i^T\right)$.


\subsection{Manifold Alignment Formulation}
The manifold alignment problem for a source data set $\mathcal{X}$ (consisting of $X$ points in $\mathds{R}^h$) and destination data set $\mathcal{Y}$ (consisting of $Y$ points in $\mathds{R}^h$) is expressed as:
\begin{align}\label{eq:manifold-alignment}
 \arg\min_{\mathbf{f},\mathbf{g}} \Bigg\{\lambda^x \sum_{i,j} [f_i - f_j]^2W^x_{ij} + \lambda^y \sum_{i,j} & [g_i - g_j]^2 W^y_{ij} \nonumber \\
&  + \mu \sum_{i\in \mathcal{P}}|f_i - g_i|^2\Bigg\}\;.
\end{align}
where $\mathbf{f} = [f_1,\dots,f_X]^T$ and $\mathbf{g} = [g_1,\dots,g_Y]^T$ are vectors in $\mathds{R}^X$ and $\mathds{R}^Y$, respectively, and $\mathcal{P}$ is the set of indices for paired points in $\mathcal{X}$\ignore{in pairwise correspondence to points in} and $\mathcal{Y}$, which are known to be at the same or close points in the lower dimensional space. Minimizing the first term guarantees that the larger $W^x_{ij}$, the smaller $f_i - f_j$, which preserves the neighborhood relations of $\mathcal{X}$ within the elements of $\mathbf{f}$. Minimizing the second term does the same in $\mathbf{g}$ for $\mathcal{Y}$. The last term in \eqref{eq:manifold-alignment} penalizes discrepancies between the paired points in the $\mathbf{f}$ and $\mathbf{g}$ vectors. $\lambda^x$, $\lambda^y$ and $\mu$ are weighting factors of the different components.

The above equation can be re-written as
\begin{equation}\label{eq:manifold-alignment-2}
\arg\min_{\mathbf{f},\mathbf{g}} \left\{\lambda^x \mathbf{f}^T L^x \mathbf{f} + \lambda^y \mathbf{g}^T L^y \mathbf{g} + \mu \left(\mathbf{f}-\mathbf{g}\right)^T\left(\mathbf{f}-\mathbf{g}\right)\ignore{\sum_{i\in \mathcal{P}}|f_i - g_i|^2}\right\}\;.
\end{equation}
where $L^x = [L^x_{ij}]$ $\forall~i,j\in\mathcal{X}$, such that:
\begin{equation}\label{eq:Laplacian}
L^x_{ij} =
\begin{cases}
\sum_{j} W^x_{ij} & \qquad i=j \\
- W^x_{ij} & \qquad j\in\mathcal{N}_i\\
0 &\qquad \mbox{Otherwise}\;,
\end{cases}
\end{equation}
and $L^y = [L^y_{ij}]$ $\forall~i,j\in\mathcal{Y}$ such that $L^y_{ij}$ is defined by replacing each $W_{ij}^x$ by $W_{ij}^y$ in \eqref{eq:Laplacian}. This problem is ill-defined. However, if a hard constraint is to be imposed so that $f_i = g_i$ $\forall~i\in P$ (i.e. as $\mu \rightarrow \infty$), and defining $\mathcal{Q}^x = \mathcal{X}\setminus\mathcal{P}$ and $\mathcal{Q}^y = \mathcal{Y}\setminus\mathcal{P}$, the problem in \eqref{eq:manifold-alignment-2} can be easily transformed into an eigenvalue problem as:
\begin{align}
&\arg\min_{\mathbf{h}}  \left\{ \frac{\mathbf{h}^T L^z \mathbf{h}}{\mathbf{h}^T \mathbf{h}} \right\} \\
&\quad \mbox{s.t.} \quad  \mathbf{h}^T\mathbf{1} = 0
\end{align}
where
\begin{equation}
\mathbf{h} = \begin{bmatrix}
\mathbf{f}_\mathcal{P} = \mathbf{g}_\mathcal{P} \\
\mathbf{f}_{\mathcal{Q}^x}\\
\mathbf{g}_{\mathcal{Q}^y}
\end{bmatrix}
\end{equation}
\begin{equation}\label{eq:laplacian_big}
L^z = \begin{bmatrix}
\lambda^x L^x_{\mathcal{P}\mathcal{P}} + \lambda^y L^y_{\mathcal{P}\mathcal{P}} & \lambda^x L^x_{\mathcal{P}\mathcal{Q}^x}  & \lambda^y L^y_{\mathcal{P}\mathcal{Q}^y}\\
\lambda^x L^x_{\mathcal{Q}^x\mathcal{P}} & \lambda^x L^x_{\mathcal{Q}^x\mathcal{Q}^x} & \mathbf{0} \\
\lambda^y L^y_{\mathcal{Q}^y\mathcal{P}} & \mathbf{0} & \lambda^y L^y_{\mathcal{Q}^y\mathcal{Q}^y}
\end{bmatrix}\;,
\end{equation}
where $L^x_{\mathcal{I}\mathcal{J}}$ $\left(L^y_{\mathcal{I}\mathcal{J}}\right)$ is a sub-matrix of matrix $L^x_{\mathcal{I}\mathcal{J}}$ $\left(L^y_{\mathcal{I}\mathcal{J}}\right)$ consisting of its entries at the intersection of the rows indexed by the elements in $\mathcal{I}$ and the columns indexed by the elements of $\mathcal{J}$. The solution to the problem is the eigenvector $\mathbf{h}$ corresponding to the smallest non-zero eigenvalue of $L^z$. According to the structure of $L^z$, $\mathbf{h}$ is structured such that it starts with the $\mathcal{P}$ aligned elements of $\mathbf{f}$ and $\mathbf{g}$, followed by the remaining data points of $\mathbf{f}$, and then ends with the remaining data points of $\mathbf{g}$.

Now, since we require an $l$ dimensional embedding ($l < h$) for the data sets, this embedding will consist of the $l$ eigenvectors $[\mathbf{h}^{(1)},\dots,\mathbf{h}^{(l)}]$ corresponding to the $l$ smallest non-zero eigenvalues of $L^z$. The final structure of the embedding $\mathcal{E}$ will be:
\begin{equation}
\mathcal{E} = \begin{bmatrix}
\mathbf{f}^{(1)}_\mathcal{P} & \mathbf{f}^{(2)}_\mathcal{P} & \dots & \mathbf{f}^{(l)}_\mathcal{P}\\
\mathbf{f}^{(1)}_{\mathcal{Q}^x} & \mathbf{f}^{(2)}_{\mathcal{Q}^x} & \dots & \mathbf{f}^{(l)}_{\mathcal{Q}^x}\\
\mathbf{g}^{(1)}_{\mathcal{Q}^y} & \mathbf{g}^{(2)}_{\mathcal{Q}^y} & \dots & \mathbf{g}^{(l)}_{\mathcal{Q}^y}
\end{bmatrix}
\end{equation}

\ignore{
Defining $h = \left[f^T|g^T\right]^T$, we can restate the problem as:
\begin{equation}
\arg\min_{h} H'(h) = \frac{h^T L^z h}{h^T h} \qquad \mbox{s.t.} \quad h^T\mathbf{1} = 0
\end{equation}
where
\begin{equation}
L^z = \begin{bmatrix}
\lambda^x L^x + U^x & -U^{xy}\\
-U^{yx} & \lambda^y L^y + U^y
\end{bmatrix}
\end{equation}
such that $U^x$, $U^y$, $U^{xy}$ and $U^{yx}$ are matrices of matching dimensions that have non-zero entries only on the diagonal:
\begin{equation}
U_{ij} = \begin{cases}
\mu \qquad & i=j \in P\\
0 \qquad & O.W.
\end{cases}
\end{equation}

If hard constraints are to be imposed so that $f_i = g_i$ $\forall~i\in P$ (i.e. as $\mu \rightarrow \infty$), we can restate the problem as an eigenvalue problem:
\begin{equation}
\arg\min_{h} C(h) = \frac{h^T L^z h}{h^T h} \qquad \mbox{s.t.} \quad h^T\mathbf{1} = 0
\end{equation}
where
\begin{equation}
h = \begin{bmatrix}
f_\mathcal{P} = g_\mathcal{P} \\
f_\mathcal{Q}\\
g_\mathcal{Q}
\end{bmatrix}
\end{equation}
\begin{equation}\label{eq:laplacian_big}
L^z = \begin{bmatrix}
\lambda^x L^x_{\mathcal{P}\mathcal{P}} + \lambda^y L^y_{\mathcal{P}\mathcal{P}} & \lambda^x L^x_{\mathcal{P}\mathcal{Q}}  & \lambda^y L^y_{\mathcal{P}\mathcal{Q}}\\
\lambda^x L^x_{\mathcal{Q}\mathcal{P}} & \lambda^x L^x_{\mathcal{Q}\mathcal{Q}} & \mathbf{0} \\
\lambda^y L^y_{\mathcal{Q}\mathcal{P}} & \mathbf{0} & \lambda^y L^y_{\mathcal{Q}\mathcal{Q}}
\end{bmatrix}
\end{equation}
}


\section{Proposed Solution using Simulated Radio Map} \label{sec:SRM-algorithm}

A simulated radio map is a model-based generated map, estimating the RSS readings from any group of wireless transmitters in any desired point or group of points in outdoor or indoor environments. Usually, such simulator are used in the radio coverage studies for network planning. Many radio map simulators have been developed in the industry using different types and details of radio propagation models. These simulators require details of the environment under study (e.g. topography, building shapes, material and dimensions in outdoor environments, and floor plans, wall material and furniture in indoor environments) as well as the positions and heights of the wireless transmitters.

In this section, we will introduce our manifold alignment based solution to directly localize users using the simulated radio map and limited deployment calibration fingerprints. Our target is to transfer the spatial correlation between neighboring simulated RSS values to the concatenation of a limited number of calibration fingerprints and one or more RSS observations with unknown locations to directly localize these observations (and thus the users observing them). The simulated radio maps indeed reflect, to a good extent, the radio propagation effects in space, such as power decay, reflections, diffractions and fading. However, due to model inaccuracies, some non-neighboring positions may end-up having quite similar simulated RSSs and thus the simulated radio maps usually suffer from neighborhood correlation outliers.

Despite this outlier problem, the two assumptions of manifold alignment generally fit these two data sets. For both the simulated radio map and the concatenated set of fingerprints and RSS observations, the nearby locations usually have more similar RSS values than those that are far away. Moreover, the two data sets are based on a common physical space (i.e. the same coordinates on the floor plan) and thus indeed have a common lower dimensional correlation. This makes the transfer learning with manifold alignment feasible.

\subsection{Offline Deployment Phase} \label{SRM-offline}
In the deployment phase, we perform the following steps:
\begin{enumerate}
\item Collect environment information (floor plan, wall thickness and materials, ...) from its building CAD files as well as the positions and heights of APs.
\item Insert this information to the radio propagation simulator to generate the simulated radio map $\mathcal{S} = \left[\left(\mathbf{s}^{(1)},p^{(1)}\right),\dots, \left(\mathbf{s}^{(S)},p^{(S)}\right)\right]$ at all the $S$ grid points of the indoor environment:
\begin{itemize}
\item $\mathbf{s}^{(i)} = \left[s^{(i)}_1,...s^{(i)}_K\right]^T$: is the simulated RSS vector from the $K$ APs of the environment at the i-th position.
\item $\mathbf{p}^{(i)} = \left[x^{(i)},y^{(i)}\right]$: $x-y$ coordinates of the i-th position.
\end{itemize}.
\item Using the simulated radio map as the source data set, compute the LLE weights $W_{ij}^x$ from \eqref{eq:weights-solution} and $L^x$ from \eqref{eq:Laplacian}. The complexity of this step is $O(S^3)$.
\item Determine a limited number $C$ of calibration locations and collect fingerprints at them. This results in the set $\mathcal{C} = \left[\left(\mathbf{c}^{(1)},p_c^{(1)}\right)..., \left(\mathbf{c}^{(C)},p_c^{(C)}\right)\right]$ of calibration RSS fingerprints:
\begin{itemize}
\item $\mathbf{c}^{(i)} = \left[c^{(i)}_1,...c^{(i)}_K\right]^T$: is the calibration RSS vector from the $K$ APs at the i-th position.
\item $\mathbf{p}_c^{(i)} = \left[x_c^{(i)},y_c^{(i)}\right]$: $x-y$ coordinates of the i-th calibration position.
\end{itemize}
\end{enumerate}
Note that calibration positions are a subset of all the positions in $\mathcal{S}$. We call these positions and their RSSs from both simulation and calibration fingerprints as paired data points.

\subsection{Server-based Online Localization Phase} \label{sec:SRM-online}
In the online localization phase, the localization server performs the following operations:
\begin{enumerate}
\item The server receives $O$ online RSS observations for localization requests $\mathcal{O} = \left[\mathbf{o}^{(1)}...,\mathbf{o}^{(O)}\right]$. These observations are obtained from $O$ localization requests from stationary users.
\item Define the following sets:
\begin{itemize}
\item $\mathcal{P}$: Indices of paired data points between $\mathcal{S}$ and $\mathcal{C}$.
\item  $\mathcal{Q}^x$: Indices of positions in $\mathcal{S}$ that are not paired with $\mathcal{C}$.
\item $\mathcal{X} = \left[\bigcup_{i\in \mathcal{P}}\mathbf{s^{(i)}}|\bigcup_{j\in \mathcal{Q}^x}\mathbf{s^{(j)}}\right]$: Simulated radio map vectors, re-arranged so that the RSSs of paired positions are brought up.
\item $\mathcal{Y} = \left[\mathbf{c}^{(1)},\dots,\mathbf{c}^{(C)}|\mathbf{o}^{(1)},\dots,\mathbf{o}^{(O)}\right]$: Concatenation of offline calibration fingerprints and online observation RSS vectors.
\item $\mathcal{Q}^y$: Indices of data points in $\mathcal{Y}$ corresponding to user observations for localization requests.
\end{itemize}
\fref{fig:datasets} depict the structure transformation from the problem's inputs and outputs to the source and destination data sets after the above ordering and concatenations.
\begin{figure}[t]
\centering
  \includegraphics[width=0.7\linewidth]{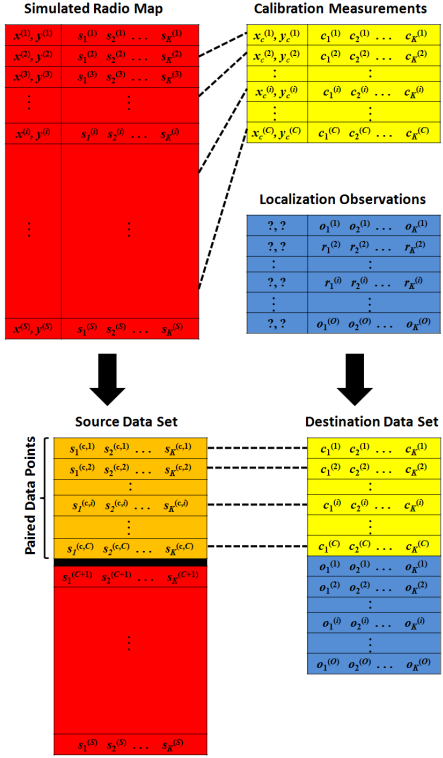}\\
  \caption{Structure of source and destination data sets for the simulated radio map algorithm}\label{fig:datasets}
\end{figure}
\item Compute the destination set neighborhood weights $W_{ij}^y$ \eqref{eq:weights-solution} and its Laplacian $L^y$ using \eqref{eq:Laplacian}. The complexity of this set is $O\left((C+O)^3\right)$.
\item Compute $L^z$ as in \eqref{eq:laplacian_big}, using:
\begin{equation}
\lambda^x = \frac{C + O}{S+C+O}  \qquad \lambda^y = \frac{S}{S+C+O}
\end{equation}
\item Calculate the eigenvalues of $L^z$, construct an $(S + O) \times l$ matrix of the $l$ eigenvectors  (each of dimension $(S+O)\times 1$) corresponding to the smallest $l$-nonzero eigenvalues. The embedding is structured as shown in \fref{fig:embedding}
\begin{figure}[t]\centering
  \includegraphics[width=1\linewidth]{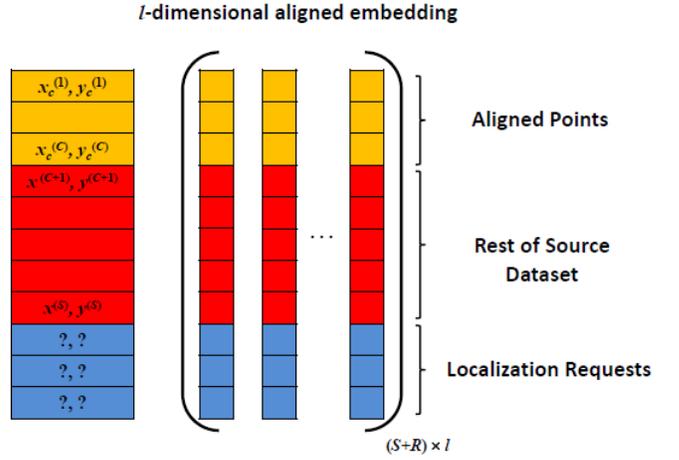}\\
  \caption{Structure of the aligned embedding $\mathcal{E}$.}\label{fig:embedding}
\end{figure}
\begin{itemize}
\item The first $C$ rows $\mathcal{E}_\mathcal{P} = [\mathbf{f}^{(1)}_{\mathcal{P}},\dots,\mathbf{f}^{(l)}_{\mathcal{P}}]$ correspond to the aligned embedding of the $C$ extended spatial vectors with the $C$ calibration fingerprints.
\item The following $(S-C)$ rows $\mathcal{E}_{\mathcal{Q}^x} = [\mathbf{f}^{(1)}_{\mathcal{Q}^x},\dots,\mathbf{f}^{(l)}_{\mathcal{Q}^x}]$ represent the embedding of remaining spatial vectors that are not paired with calibration fingerprints.
\item The last $O$ rows $\mathcal{E}_{\mathcal{O}} = [\mathbf{g}^{(1)}_{\mathcal{Q}^y},\dots,\mathbf{g}^{(l)}_{\mathcal{Q}^y}]$ represent the embedding of the users observations for localization.
\end{itemize}
\item For each row of $\mathcal{E}_\mathcal{O}$:
\begin{itemize}
\item Compute its distance to all the rows of $\mathcal{E}_\mathcal{P}$ and $\mathcal{E}_{\mathcal{Q}^x}$.
\item Attach the position of the nearest row to this observation and forward it to the requesting user.
\end{itemize}
\end{enumerate}
The overall complexity of the algorithm is $O((C+O)^3 + (S+O)^3)$.

\ignore{Note that the same algorithm can be implemented in the client side (i.e. the user smartphone). In this case, the client's software needs to download from the server $L^x$ (a compact $S\times S$ matrix) and the calibration fingerprints $\mathcal{C}$ (a small data set).}

\subsection{Algorithm Modifications for Walking Users}
Naturally, a walking user requesting localization can send $U$ subsequent RSS observations, which will definitely represent locations of close-by coordinates. This information can be used in smoothing out positions of each group of subsequent observations. This can be done in two places in the above algorithm:
\begin{itemize}
\item In step 3, enforce neighborhood and high weights for each group of subsequent observations.
\item After step 6, replace each outlier (position that is far away from its immediate previous and following estimated locations) by the centroid of its immediate predecessor and successor estimated locations.
\end{itemize}

\subsection{Device-based Online Localization Phase}
The same algorithm introduced in the previous section can be implemented in the device side (e.g. the user smartphone). In this case, the device's localization software needs to perform a one-time download of $L^x$ (a compact $S\times S$ matrix) and the calibration fingerprints $\mathcal{C}$ (a small data set) from the server. The number of localization requests will be only one (i.e. $O=1$), which corresponds to the device's own RSS reading. With this data available at the device, it can then apply steps 3 to 6 described in the previous section to find its location. The complexity of this process is $O(S^3 + C^3)$, which is quite affordable on mobile, battery empowered devices.

Note that, in case of moving user, the localization frequency on the mobile device can be reduced by incorporating the information from the device sensors, such as accelerometer, gyroscope and compass, to the localization software. Indeed, these sensors can track the user movement, and thus its position, within some predetermined time interval after the latest run of the RSS-based localization algorithm. Consequently, the device will only need to run the RSS-based localization algorithm at the beginning of these time intervals to correct any drifts obtained from tracking.\ignore{ Clearly, this will reduce the frequency of running the RSS-based localization algorithm, thus saving on overall complexity and battery consumption.}

\subsection{Radio Map Construction}
To construct the radio map using the above algorithm, the server builds an observation directory $\mathfrak{D}(p)$ for each of the $S$ plan coordinates, having $N$ entries, each of size $K\times 1$, to store observations. Each localization observation, sent to the server and determined by the algorithm to be at location $p^*$, is stored in an entry of the directory entry $\mathfrak{D}(p^*)$.

When the number of accumulated observations at a give position reaches the required limit $N$, the radio map construction algorithm computes the average of the $N$ readings in $\mathfrak{D}(p)$, and sets this average as the estimated RSS reading at point $p$. When the RSSs of all the points in $\mathcal{S}$ are estimated, the algorithm stops and declares the estimated radio map. Clearly, this method makes the radio map construction much simpler as it is built through casual walks in the environment after performing a few precise calibration measurements at known positions.

\section{Proposed Solution using Plan Coordinates} \label{sec:PC-algorithm}
In this section, we will modify the previous algorithm to employ the plan coordinates information instead of the simulated radio map in the direct localization of stationary users. The target is to transfer the spatial correlation of physically neighboring points to the set of calibration and online observation measurements to directly localize the users. Although it does not reflect all the aspects of radio propagation, the plan coordinates emphasize perfect physical neighborhood relations in the source data set. Also, the plan coordinates requires much less details about the environment, especially the location and heights of APs that are usually difficult to obtain in indoor environments.

Again, the two assumptions of manifold alignment perfectly fit these two data sets. The nearby coordinates (calibration and observation data points) usually have smaller distances (more similar RSS values) than those that are far away. Moreover, the two data sets are based on the common floor plan physical space and thus indeed have a common lower dimensional correlation, which makes the transfer learning with manifold alignment feasible.

\subsection{Offline Deployment Phase} \label{sec:PC-description}

In the deployment phase, we perform the following steps:
\begin{enumerate}
\item Collect environment information (floor plan, wall thickness and materials, ...) from its building CAD files.
\item Set up a grid point system on the floor plan and determine its coordinates. The result is the source data set $\mathcal{S} = \left[\left(p^{(1)}\right),\dots, p^{(S)}\right]$, where $\mathbf{p}^{(i)} = \left[x^{(i)},y^{(i)}\right]$ is $x-y$ coordinates of the i-th position.
\item In case of very thick or metallic walls, dissociate any two neighboring points that are located at opposite sides of these walls. This dissociation enforces that these points do not become neighbors when computing the source weights.
\item Repeat the steps 3 and 4 in \sref{SRM-offline}
\end{enumerate}
Clearly, this approach reduces the deployment load compared to the simulated radio map approach as it removes the need to know the positions of access points and to run simulations to obtain the source data set.

\subsection{Online Localization Phase}
The online localization phase of this approach is similar to that in \sref{sec:SRM-online} after adding the following definition and making the following adjustments.
\begin{itemize}
\item Define $\mathbf{\hat{p}^{(i)}}$: $1\times K$ extended coordinate vector, in which the x and y elements of $\mathbf{p^{(i)}}$ are alternatively repeated inside the vector $\mathbf{\hat{p}^{(i)}}$ until its number of elements becomes equal to $K$. These vectors are used instead of the original coordinate vectors to match the dimensions of the calibration data without any loss in the actual distances between points.
\item Modify the definition of set $\mathcal{X} = \left[\bigcup_{i\in P}\mathbf{\hat{p}^{(i)}}|\bigcup_{j\in \mathcal{Q}^x}\mathbf{\hat{p}^{(j)}}\right]$ to be the extended coordinate vectors, re-arranged so that the coordinates of paired positions are brought up.
\end{itemize}
\fref{fig:datasets} depict the structures of the source and destination data sets after the above ordering and concatenations.
\begin{figure}[t]
\centering
  \includegraphics[width=0.7\linewidth]{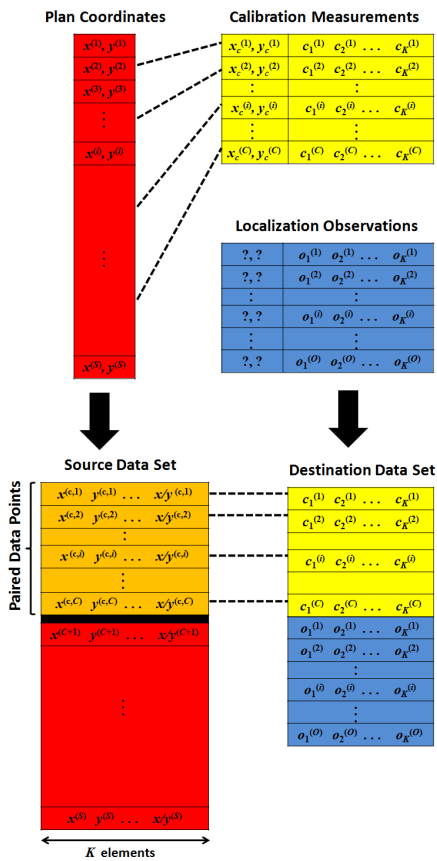}\\
  \caption{Structure of source and destination data sets for the plan coordinates algorithm}\label{fig:datasets}
\end{figure}
The rest of the algorithm runs in the exact same way. Also, the same modifications for device based localization and walking users can be applied to this approach.

\section{Performance Testing of Localization Solutions} \label{sec:sim-L}
To test the proposed solutions, we considered two indoor environments:
\begin{enumerate}
\item Bahen Center, University of Toronto, the 4th floor.
\item Siradel's building in Rennes, France.
\end{enumerate}

\subsection{Bahen Center}
\begin{figure}[t]
\centering
\includegraphics[width=1\linewidth]{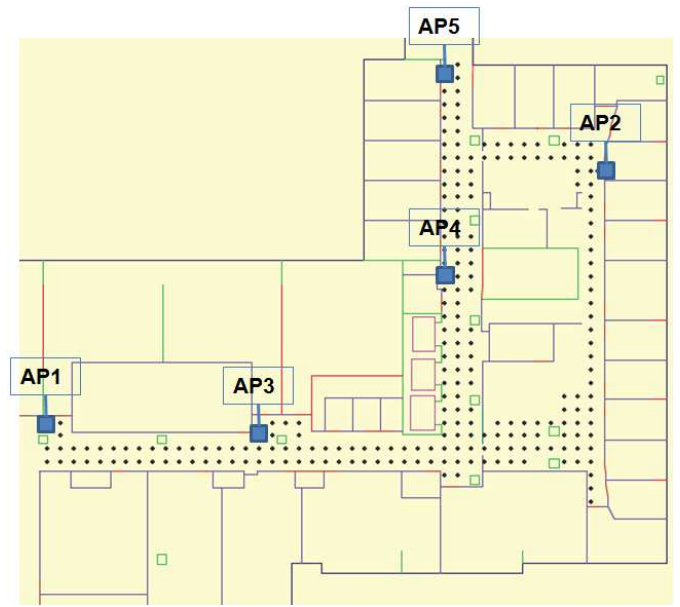}
\caption{Bahen center floor plan illustrating APs and data collection/simulation grid points}\label{fig:bahen-plan}
\end{figure}

The first testing environment is the 4th floor of Bahen Center, University of Toronto. The environment floor plan spans an area of 40m $\times$ 30m and is depicted in \fref{fig:bahen-plan}. We deployed 5 linksys APs at the shown locations in the figure, in order to be able to compare our results to the ones obtained in \cite{6042868} for the same environment and APs. Nonetheless, the performance of our algorithm would have not been affected if we selected any pre-installed APs from the environment with known locations. The study area is divided into 219 grid points at which RSS data is both measured and simulated. The spacing between any two grid points is 1m. Note that the mean localization error obtained by a very sophisticated and recent localizer, using full calibration, location clustering, and best 5 AP selection (compared to only static 5 APs in our case) was reported to be 2m \cite{6042868}. The performance of this localizer will be the reference comparison scheme, to which we will compare our results.

After setting up the coordinate system shown in \fref{fig:bahen-plan}, the neighborhood weights for the spatial correlation approach can be computed once (as long as there is no drastic floor plan change) and are saved in the database for direct use in the manifold alignment localization process.\\

\subsubsection{Generation of Simulated using Volcano Lab} \label{sec:model}$\quad$\\
Volcano Lab (VLAB) \cite{VLAB} is a platform desinged by Siradel France, which has various capabilities of simulating indoor and outdoor radio coverage of wireless transmitters. It can run both simple propagation and ray-tracing algorithms to estimate the RSSs at any resolution. To simulate the radio map of indoor environments, it simply requires the floor plan information (e.g. from its CAD file) and the positions and heights of APs.

In our environment, we ran VLAB for the APs in \fref{fig:bahen-plan} using 4 different simulation setups.
\begin{itemize}
\item Direct path model: Involves only the direct path in the estimation of the RSS reading at each point.
\item 1-Reflection model: Involves the direct path and one reflection in the estimation of the RSS reading at each point.
\item 2-Reflection model: Involves the direct path and two reflections in the estimation of the RSS reading at each point.
\item Ray-tracing model: Involves 4 reflections and one diffraction of each propagation ray into the estimation of RSSs at each point.
\end{itemize}
Clearly, the further we go down in the above list, we achieve a higher fidelity of the RSS estimation but at the same time we increase the simulation complexity.

After obtaining the simulated RSS map, the neighborhood weights for the simulated radio map approach are computed only once (as long as there is no floor plan change) and saved in the database for direct use in the manifold alignment localization process.\\

\subsubsection{Data Collection and Testing Setup}\label{sec:data-collection}$\quad$\\
We performed a full measurement campaign, in which we took RSS fingerprints at all grid points depicted in \fref{fig:bahen-plan}, using an RSS collecting software loaded on an HP IPAQ device. Although we do not need all this data in the actual implementation of the algorithm, we collected all fingerprints to test the algorithm's performance for different subsets of calibration measurement positions and to make sure it is insensitive to the chosen calibration points.

In the offline phase, we built many calibration data sets, each of which consisting of a given percentage of the collected fingerprints scattered over the designated area. In the online phase, the algorithm receives $O$ localization observations (from either $O$ stationary users or one walking user) and estimates their positions. A large number of localization tests were performed for different $O$ localization readings and using different calibration data sets. The following figures depict the mean localization errors averaged over a large number of localization tests for each setup. All percentages of calibration fingerprints are normalized against the total number of grid points (219 points).\\

\subsubsection{Effect of Neighborhood Size}
\begin{figure}[t]
\centering
  \includegraphics[width=1\linewidth]{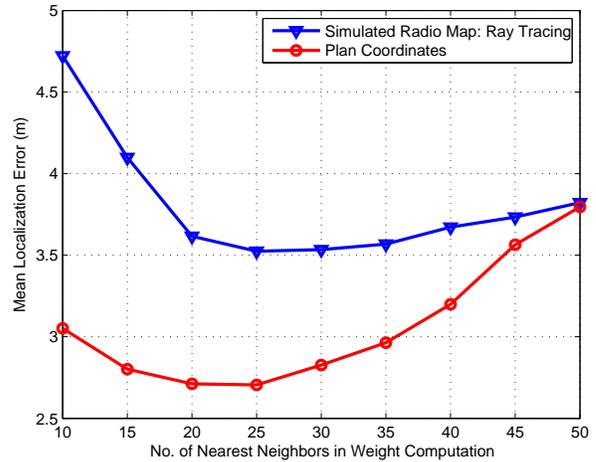}
  \caption{Bahen center results: Mean localization error performance against the number of neighbors selected in the weigh computations}\label{fig:N}
\end{figure}

$\quad$\\
We first test the effect of changing the number of neighbors in the neighborhood weight computation process on the performance of the algorithm. \fref{fig:N} depicts the mean localization error performance of both stationary and walking user algorithms, against the number of neighbors in the neighborhood weight computation process. The percentages of calibration fingerprints and localization observations are 25$\%$ and 5$\%$, respectively.
The figure clearly shows that, for both simulated radio map and plan coordinate algorithms, the number of neighbors in weight calculations is a significant factor in determining the localization error level. If the number of neighbors is small (10-15), the performance level is worse than when the number of neighbors is larger (in the range of 20 - 25 neighbors, i.e. 10-12$\%$ of the total number of grid points). This can be explained by the stronger effect of outliers in weight computations, when the number of neighbors is small. For the case of simulated radio map, if there exist several points, which are practically far from the point for which we compute the weights, but have very close RSS vectors to it, the weight computation would be more affected by these outlier points if the total number of neighbors is smaller. The larger the number of neighbors, the smaller the percentage of these outliers, the smaller their effect in misrepresenting points in the lower dimensional space. For the plan coordinate case, these outliers may result from the misrepresented fading instances and thus the same effect occurs.

However, we can see that for larger numbers of neighbors (35 - 50 neighbors), the performance degrades again. This can be interpreted by the concept of loose neighborhoods. If the number of neighbors has gone too high, we are mainly relating each point to a lot of points that are not in its vicinity and thus the concept of neighborhood dilutes and results in the shown performance degradation.\\

\subsubsection{Effect of Calibration Fingerprinting Load}
\begin{figure}[t]
\centering
  \includegraphics[width=1\linewidth]{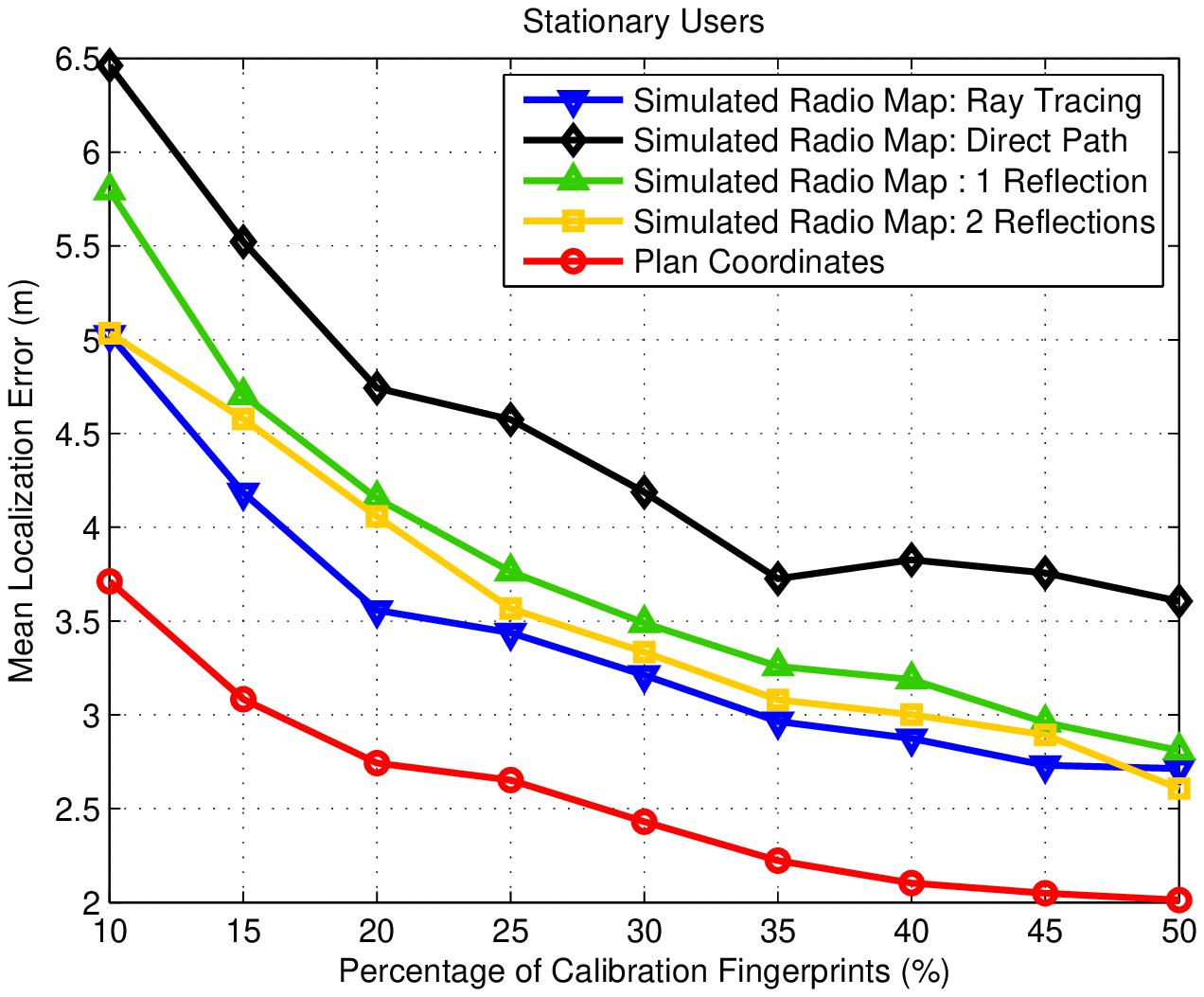}
  \caption{Bahen center results: Mean localization error performance against the percentage of calibration fingerprints}\label{fig:L-Bahen-Stationary}
\end{figure}

$\quad$\\
\fref{fig:L-Bahen-Stationary} depicts the mean error of localizing 11 observations using the proposed simulated radio map and plan coordinates algorithms against the percentage of the calibration fingerprints. The number of neighbors in the weight computation process is set to 25 (11$\%$ of the total data set size).
As expected, the figure shows that the larger the percentage of calibration fingerprints, the better the localization error. We can also see that the plan coordinates approach achieves a better performance than all the simulated ratio maps. This is explained by the fact that the plan coordinates approach preserves perfect physical neighborhoods where as all simulated maps have noisy inaccuracies in their RSS estimations, which degrades their neighborhood weight computations. It also shows that neighborhood correlation (represented in the plan coordinates) matters more in determining more accurate localization results compared to propagation effects (represented in the simulated radio maps). This is even more apparent within the different simulated maps as it is clear that, the more details involved in the RSS estimation, the lower the localization error. However, this comes at the expense of simulation complexity. Nonetheless, the plan coordinates approach avoids this complexity completely, while achieving an even better result.

\begin{figure}[t]
\centering
  \includegraphics[width=1\linewidth]{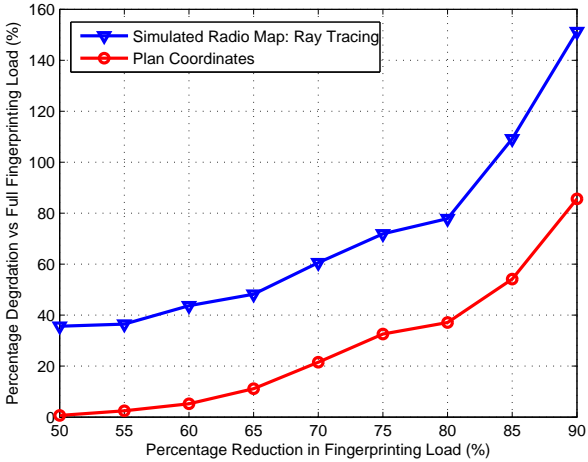}
  \caption{Bahen center results: Percentage degradation in mean localization error performance, compared to the reference full-calibration localizer}\label{fig:LI-Bahen-Stationary}
\end{figure}

\fref{fig:LI-Bahen-Stationary} depicts the percentage degradation in mean localization error performance, compared to the reference full-calibration localizer in \cite{6042868}, against the percentage reduction in the fingerprinting load for the same parameters as in \fref{fig:L-Bahen-Stationary}
We can see from this figure that, our proposed scheme using plan coordinates can achieve a degradation in localization error of less than 20$\%$  (i.e 0.4m) for as high as 70$\%$ reduction in the calibration load, and a degradation of less than 40$\%$ (i.e 0.8m) for as high as 80$\%$ reduction in the calibration load. Moreover, this plan coordinates approach does not require the knowledge of the positions of access points and avoids the need for simulations and other pre-processing loads to obtain the source data set.

\begin{figure}[t]
\centering
  \includegraphics[width=1\linewidth]{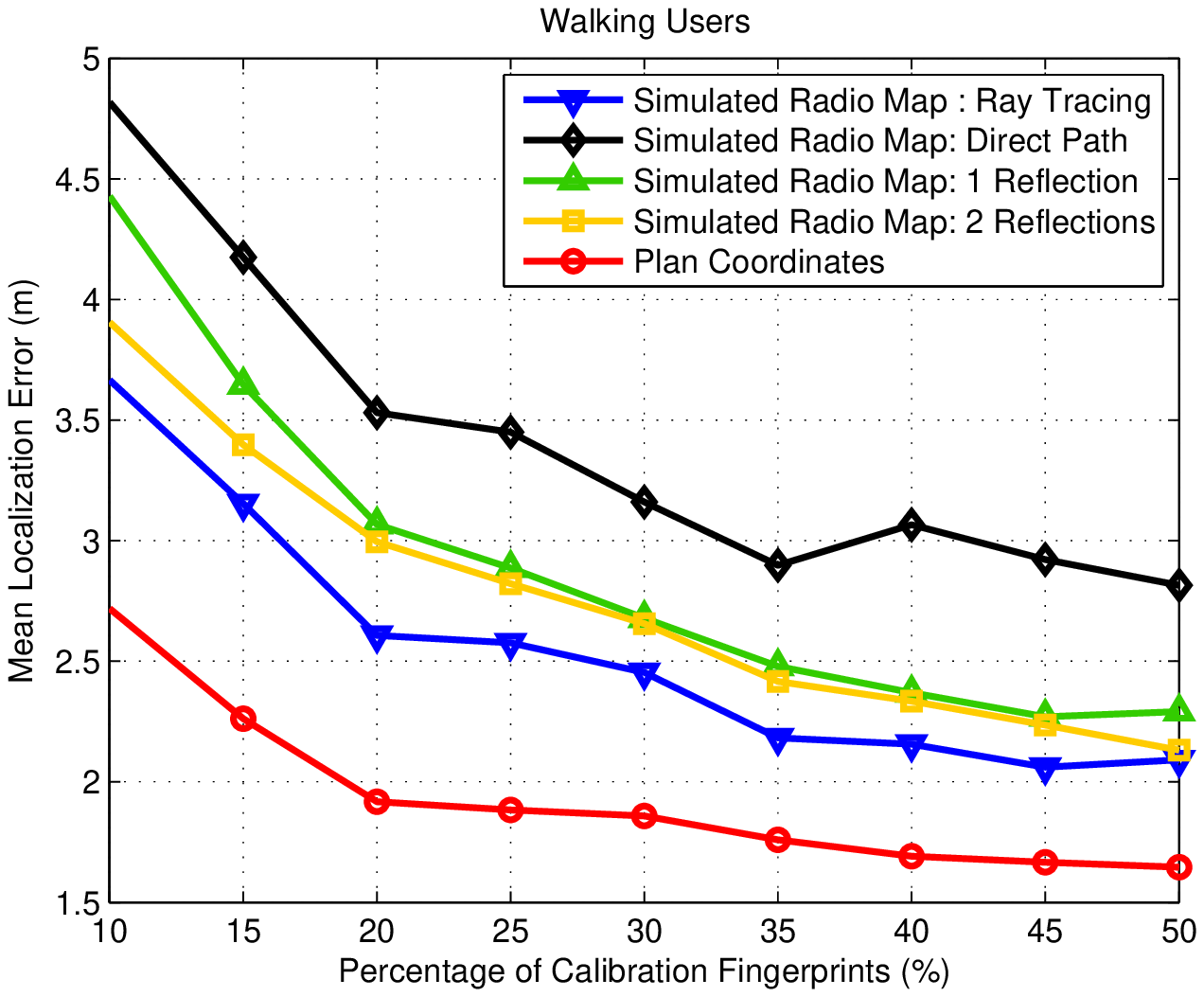}
  \caption{Bahen center results: Mean localization error performance against the percentage of calibration fingerprints}\label{fig:L-Bahen-Walking}
\end{figure}

\fref{fig:L-Bahen-Walking} depicts the mean localization error performance of our proposed solutions for walking users against the percentage of the calibration fingerprints. The number of neighbors and observations are also set to 25 and 11 points, respectively.
Again, the performance of the plan coordinates approach achieve a lower localization error compared to all simulated radio maps. Due to the exploitation of RSS correlation for walking users, the mean localization error is dropped to 2.3 to 1.8 m for 15-30$\%$ of calibration fingerprints.\\

\subsubsection{Effect of Localization Observations}
\begin{figure}[t]
\centering
  \includegraphics[width=1\linewidth]{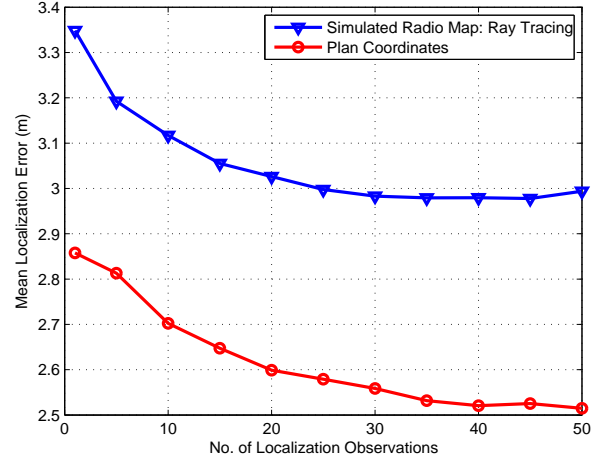}
  \caption{Bahen center results: Mean localization error performance against the percentage of localization observations}\label{fig:U}
\end{figure}

$\quad$\\
\fref{fig:U} depicts the mean localization error performance against the number of localization observations, for 25 neighbors and 25$\%$ calibration fingerprints.
Both curves show that as the number of localization observations increases, the localization error slightly decreases. Nonetheless, this change in performance from 1 observation to 50 observations is in less than 35 cm, which is not significant difference for most indoor localization applications. For stationary users, this result implies that localization can be done with quite similar error levels for small or large number of localization requests. For walking users, it implies that no delay is required to start the localization algorithm. It is enough for a user to collect 2 to 5 observations on its path to obtain its location.

\subsection{Siradel Building}
\begin{figure}[t]
\centering
\includegraphics[width=1\linewidth]{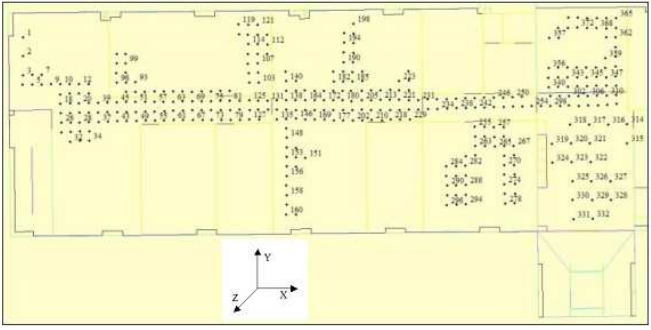}
\caption{Siradel's building floor plan illustrating the data collection/simulation grid points}\label{fig:Rennes-plan}
\end{figure}

$\quad$\\
We run another test on the collected database in Siradel Building at Rennes, France. This environment, depicted in \fref{fig:Rennes-plan}, consisted of an area of 40m $\times$ 20m, with 4 APs and 302 grid points. The spacing between grid points is variable between few tens of centimeters up to few meters.

\begin{figure}[t]
\centering
  \includegraphics[width=1\linewidth]{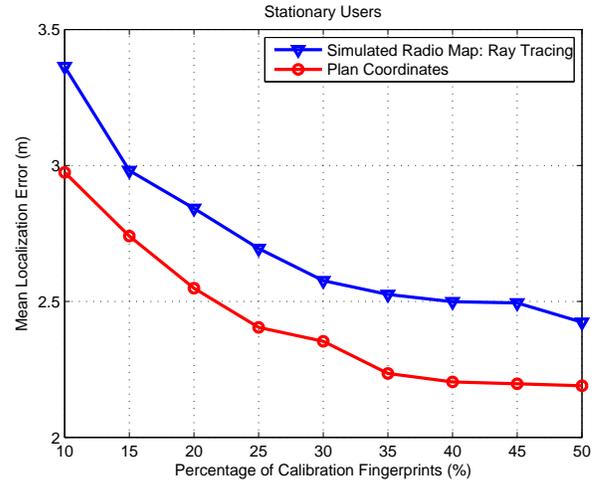}
  \caption{Siradel building results: Mean localization error performance against the percentage of calibration fingerprints}\label{fig:L-Rennes-Stationary}
\end{figure}

\begin{figure}[t]
\centering
  \includegraphics[width=1\linewidth]{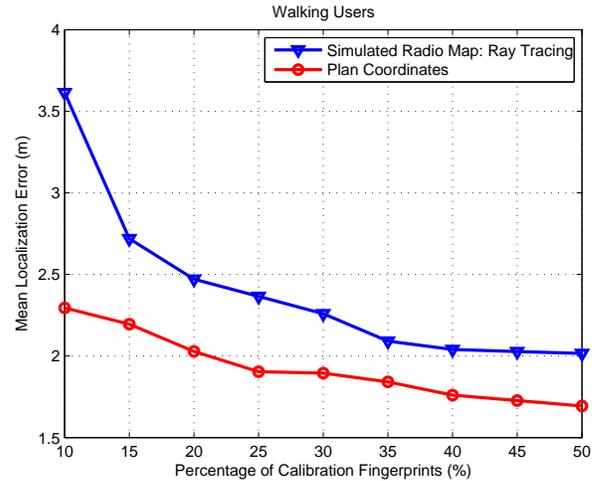}
  \caption{Mean localization error performance against the percentage of calibration fingerprints}\label{fig:L-Rennes-Walking}
\end{figure}

Figures \ref{fig:L-Rennes-Stationary} and \ref{fig:L-Rennes-Walking} depict the mean localization error performance of our proposed algorithms against the percentage of the calibration fingerprints, for stationary and walking users, respectively.

As was observed in the Bahen Center results, we can see that the plan coordinates approach achieves a better performance than all the simulated ratio maps. For as low as 15-30$\%$ of the calibration effort, the plan coordinates algorithm achieves a mean error of 2.7m - 2.4m and 2.2m - 1.9m for stationary and walking users, respectively.

\section{Performance Testing of Radio Map Construction} \label{sec:sim-R}
In this section, we test the performance of radio map construction using our proposed solution. As explained earlier, the algorithm operates in the same manner for localization, but then stores all localized reading at their estimated positions. It then declares the average of $N$ readings per location as its estimated RSS. A large number of tests was performed for different setup parameters. The following figures depict the root mean square (RMS) error of estimated points using our proposed algorithm, with respect to the actual radio map obtained from the full measurement collection explained in \sref{sec:data-collection}. They also depict the RMS improvement of the overall map (including both calibrated and estimated points) compared to the simulated radio map. The RSS estimation averaged over a large number of tests for each setup.

\begin{figure}[t]
  \centering
  \includegraphics[width=1\linewidth]{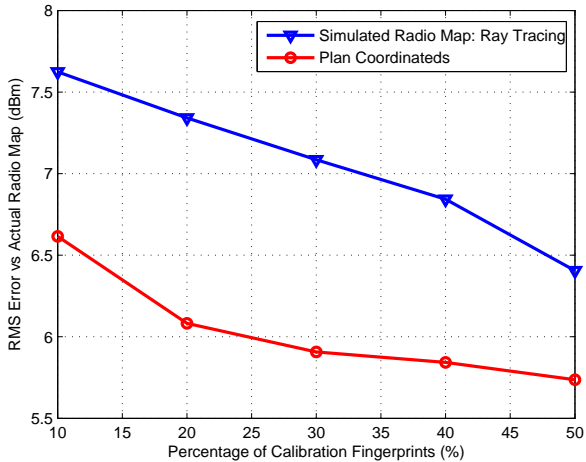}\\
  \caption{RMS error in radio map estimation for different percentages of calibration load.}\label{fig:RMSEvsL}
\end{figure}

\begin{figure}[t]
  \centering
  \includegraphics[width=1\linewidth]{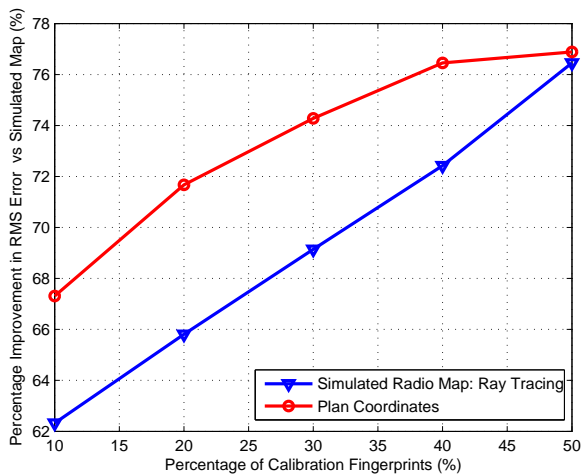}\\
  \caption{Percentage improvement in RMS error for different percentages of calibration load compared to the simulated map.}\label{fig:RMSEIvsL}
\end{figure}

\fref{fig:RMSEvsL} depicts the root mean square (RMS) error, against the percentage of calibration fingerprints. The figure compares the performance of our proposed algorithm using both the simulated radio map with ray tracing and the plan coordinates. \fref{fig:RMSEIvsL} depicts the achieved improvement in RMS error for the overall map against the same percentages of calibration load. In both figures, the number of accumulated observations per point (i.e. $N$) is 20.
As expected, both figures show that the larger the percentage of calibration load, the lower the achieved RMS error in RSS at the estimated points and the larger the achieved improvement compared to the simulated radio map. Both figures also show that the plan coordinate data set achieves a better performance compared to the simulated radio map. Clearly, this is a natural result from the ability of the plan coordinates to achieve lower localization error, which makes the accumulated readings at each location more accurate. For the plan coordinate data set, we can see in \fref{fig:RMSEvsL} that the reduction in the RMS error, when changing the calibration load from 10$\%$ to $50\%$, is only in the range of 1 dBm. Moreover, we can see from \fref{fig:RMSEIvsL} that, we can obtain 70$\%$ improvement in our knowledge of the map, compared to the simulated one, with only 16$\%$ of the full calibration load.

\begin{figure}[t]
  \centering
  \includegraphics[width=1\linewidth]{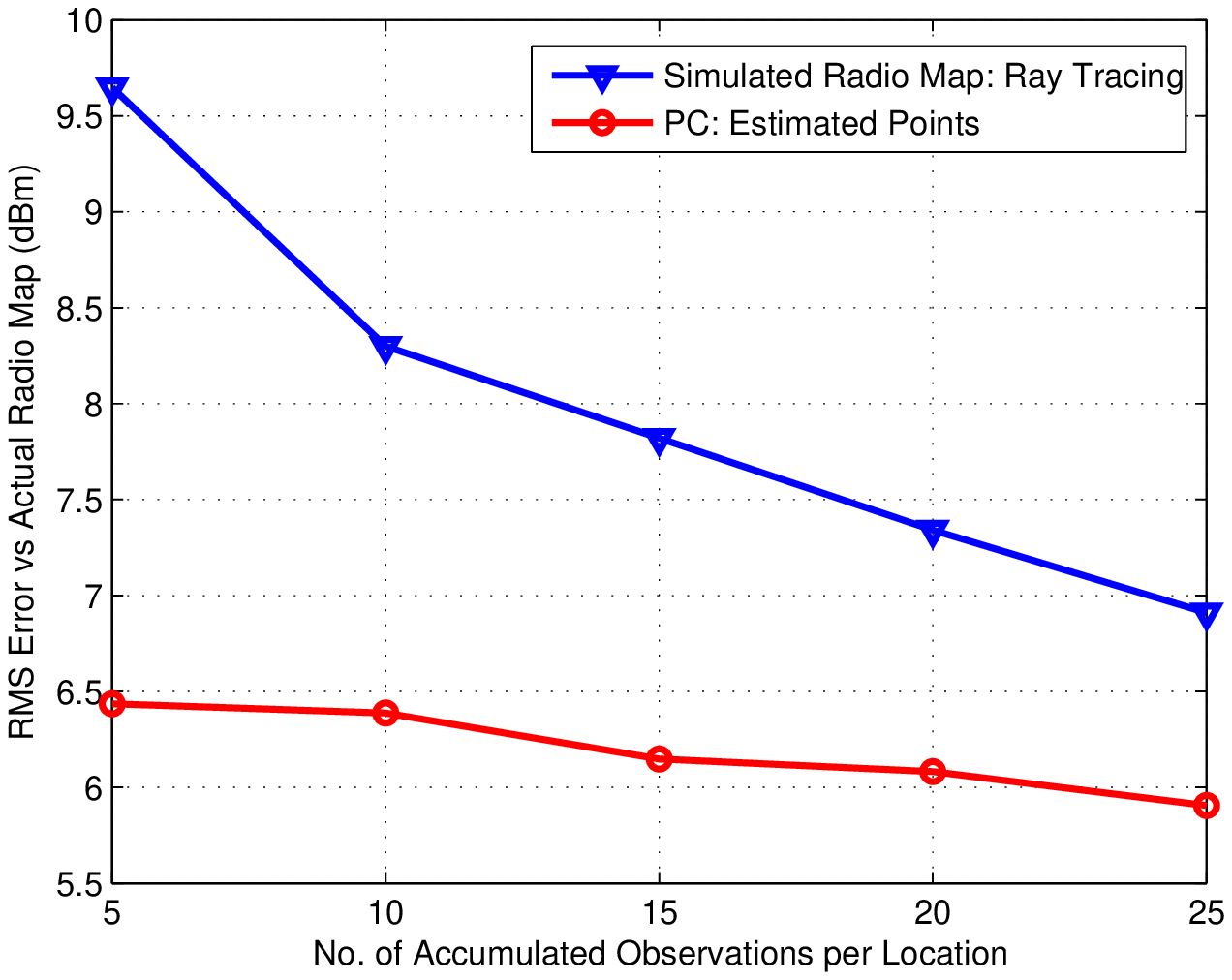}\\
  \caption{RMS error in radio map estimation for different numbers of accumulated observations per map point.}\label{fig:RMSEvsN}
\end{figure}

\begin{figure}[t]
  \centering
  \includegraphics[width=1\linewidth]{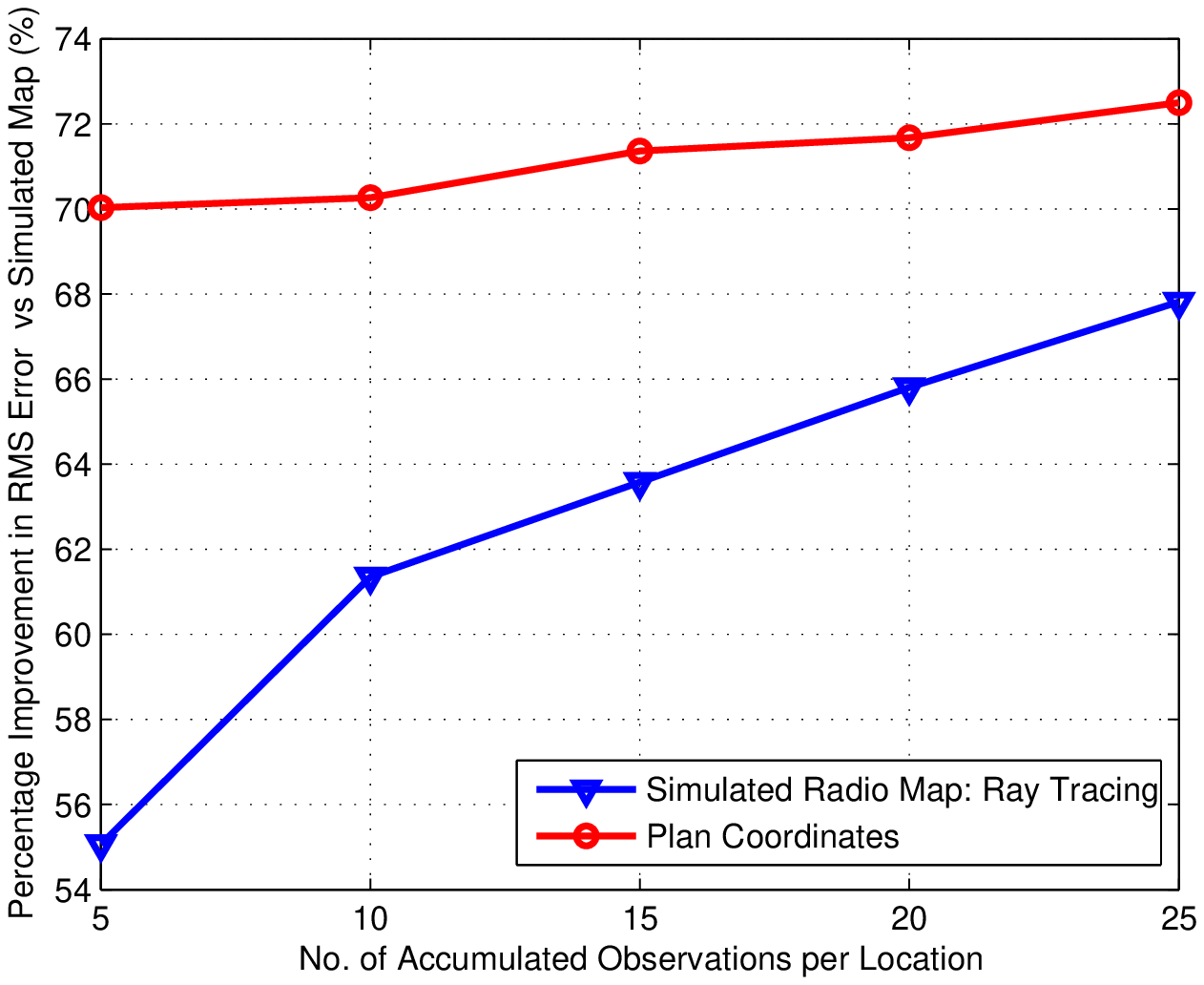}\\
  \caption{RMS error in radio map estimation for different numbers of accumulated observations per map point.}\label{fig:RMSEIvsN}
\end{figure}

\fref{fig:RMSEvsN} and \fref{fig:RMSEIvsN} depicts the effect of changing the number of accumulated observations per map location on the algorithm's performance.
We can see from the figure that this effect is minor, especially when the plan coordinates data set is used. When the number of observations is reduced from 25 to 5, the increase in the RMS error is only 0.5 dBm. Moreover, the improvement in the RMS error, compared to the simulated map, is reduced by less than 2$\%$. Thus, we can greatly reduce the required time to build the map, by collecting less observations per map point, while not significantly degrading the algorithm's performance.

\section{Conclusion} \label{sec:conclusion}
In this paper, we proposed a joint indoor localization and radio map construction scheme that can be directly deployed and employed with limited calibration load in indoor environments. The proposed scheme employs a source spatial correlation preserving data set and a limited number of calibration fingerprints. The knowledge of this source data set is transferred to the limited calibration fingerprints and the localization observations to perform direct localization using manifold alignment. By accumulating this information about localized readings, this scheme can also simultaneously construct the radio map with limited calibration. We proposed and tested correlation preserving source data sets, namely the plan coordinates and the simulated radio map. For moving users, we exploit the correlation of their reported observations to improve the localization accuracy. The online performance testing in two different indoor environments favors the use of the plan coordinates to achieve better results compared to the simulated radio maps. It also showed that, for as low as 70$\%$ to 80 $\%$ reduction in the complete fingerprinting load, our approach can achieve only a 0.4m to 0.8m increase to the full-fingerprinting localization error. The results also show that, by accumulating a few observations per location, our scheme can achieve 70$\%$ improvement in the radio map knowledge compared to the simulated one, with only 16$\%$ of the full calibration load. These gains are obtained with very limited deployment calibration and pre-processing efforts.

In the next phase of this work, we will explore the possibilities of further reduction in calibration efforts using unsupervised manifold alignment using both simulated radio maps and plan coordinate approaches. We will also investigate whether combining both the simulated radio map and plan coordinates would add further improvement and/or robustness to the algorithms at the cost of additional simulation and information collection complexities.

\bibliographystyle{IEEEtran}
\bibliography{IEEEabrv,bibfile}

\end{document}